\documentclass{article}

\usepackage[utf8]{inputenc}
\usepackage[english]{babel}
\usepackage{csquotes}
\usepackage{authblk}
\usepackage[raggedright]{titlesec}

\usepackage{graphicx}
\usepackage[usenames,dvipsnames,svgnames,table]{xcolor}
\usepackage{float}
\usepackage[a4paper,margin=1.1in,footskip=0.25in]{geometry}

\usepackage{url}
\usepackage[hyperindex,colorlinks,hyperfootnotes = false,linkcolor=blue,urlcolor=black,citecolor=blue]{hyperref}

\usepackage{bm}
\usepackage{amssymb}
\usepackage{amsmath}

\usepackage[doi=false,maxbibnames=10,maxcitenames=2,mincitenames=1,articlein=false,giveninits,sorting=none,style=ext-numeric,useprefix]{biblatex}
\usepackage{xpatch}
\xpatchbibdriver{online}
  {\printfield{entrysubtype}}
  {\printfield{entrysubtype}
   \newunit\newblock
   \printfield{addendum}}
  {}
  {}
  
\usepackage[default]{sourcesanspro}

\newbibmacro{string+doiurl}[1]{%
  \iffieldundef{doi}
    {\iffieldundef{url}
       {#1}
       {\href{\thefield{url}}{#1}}}
    {\href{\thefield{doi}}{#1}}}

\makeatletter
\def\blx@driver#1{%
  \ifcsdef{blx@bbx@#1}
    {\usebibmacro{string+doiurl}{\csuse{blx@bbx@#1}}}
    }
\makeatother  

\DefineBibliographyStrings{english}{
  byeditor = {edited by},
  editor   = {ed},
  editors  = {eds},
}

\renewcommand*{\jourvoldelim}{\addcomma\space}

\DeclareDelimFormat[bib,biblist]{editortypedelim}{\addspace}
\DeclareDelimFormat[bib,biblist]{nametitledelim}{\addcolon\space}

\DeclareFieldFormat{editortype}{\bibsentence\mkbibparens{#1}}
\DeclareFieldFormat{issuedate}{#1}
\DeclareFieldFormat{pages}{#1}
\DeclareFieldFormat{titlecase}{%
  \ifboolexpr{
    test {\ifentrytype{article}}
    or
    test {\ifentrytype{incollection}}
  }
    {\MakeSentenceCase*{#1}}
    {#1}}
\DeclareFieldFormat{titlecase:booktitle}{#1}
\DeclareFieldFormat{titlecase:journaltitle}{#1}
\DeclareFieldFormat[article,incollection]{title}{\mkbibbold{#1}}
\DeclareFieldFormat[article]{volume}{\mkbibbold{#1}}

\renewbibmacro*{journal+issuetitle}{%
  \usebibmacro{journal}%
  \setunit*{\addspace}%
  \usebibmacro{issue+date}%
  \setunit*{\jourvoldelim}%
  \iffieldundef{series}
    {}
    {\setunit*{\jourserdelim}%
     \printfield{series}%
     \setunit{\servoldelim}}%
  \usebibmacro{volume+number+eid}%
  \setunit{\addcolon\space}%
  \usebibmacro{issue}%
  \newunit}

\renewbibmacro*{begentry}{\midsentence}

\DeclareFieldFormat{labelnumberwidth}{#1\adddot}

\usepackage{xcolor}

\title{Connecting the dots in ethology: applying network theory to understand neural and animal collectives}

\makeatletter
\newcommand{\printfnsymbol}[1]{
  \textsuperscript{\@fnsymbol{#1}}
}
\makeatother

\author[1]{Adam Gosztolai\thanks{corresponding authors: \href{mailto:adam.gosztolai@epfl.ch}{adam.gosztolai@epfl.ch};  \href{mailto:pavan.ramdya@epfl.ch}{pavan.ramdya@epfl.ch}}
}
\author[1]{Pavan Ramdya\hspace{-1mm}\printfnsymbol{1}\hspace{-1.5mm}
}

\affil[1]{Neuroengineering Laboratory, Brain Mind Institute \& Interfaculty Institute of Bioengineering, EPFL, Lausanne, Switzerland}
\date{} 

\begin{document}
    
\maketitle

\begin{abstract}
A major goal shared by neuroscience and collective behavior is to understand how dynamic interactions between individual elements give rise to behaviors in populations of neurons and animals, respectively. This goal has recently become within reach thanks to techniques providing access to the connectivity and activity of neuronal ensembles as well as to behaviors among animal collectives. The next challenge using these datasets is to unravel network mechanisms generating population behaviors. This is aided by network theory, a field that studies structure-function relationships in interconnected systems. Here we review studies that have taken a network view on modern datasets to provide unique insights into individual and collective animal behaviors. Specifically, we focus on how analyzing signal propagation, controllability, symmetry, and geometry of networks can tame the complexity of collective system dynamics. These studies illustrate the potential of network theory to accelerate our understanding of behavior across ethological scales.
\end{abstract}

\section*{Highlights}

\begin{itemize}
    \item Datasets of neural connectivity and function as well as animal tracking are driving a shift towards a network-based understanding of individual and collective animal behaviors.
    \item Neuronal and animal interaction networks represent two interleaved computational layers upon which sensing, information processing, and behavior emerge.
    \item Both neural activity and animal behavior can be represented as dynamic signals over networks.
    \item A rich toolbox of network theory concepts, including signal propagation, controllability, symmetry, and network geometry can be applied to discover structure-function relationships in networks.
\end{itemize}

\section*{Introduction}

\begin{figure*}[t]
    \centering
    \includegraphics[width = \textwidth]{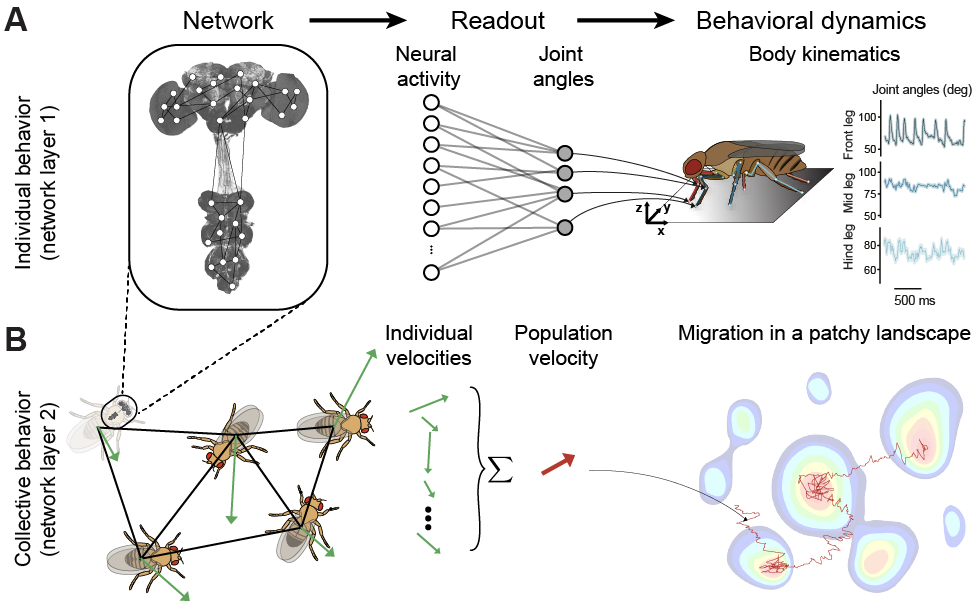}
    \caption{\textbf{Information processing in collective animal systems occurs across hierarchical layers.} \textbf{(A)} To orchestrate the actions of individual animals, populations of neurons interact, generating circuit dynamics. These dynamics converge and are read out by muscles, causing bodypart movements and ethological behaviors. \textbf{(B)} At a higher hierarchical layer, individual animals---with diverse preferences---transmit their behaviors (e.g., heading direction) as sensory signals to a social network that coordinates group dynamics, e.g., moving through a patchy nutrient landscape (adapted from \cite{Gosztolai2019}).}
    \label{figure_1}
\end{figure*}

\sloppy A central goal in neuroscience is to understand how animal behavior is orchestrated by the activity of formidably complex neuronal networks \autocite{Swanson2016}. Parallel efforts in collective animal behavior have addressed an analogous question at a larger scale of organisation: how do population-level behaviors arise from the interactions between individual animals \cite{Sumpter2006}. Both fields have, until now, favored a reductionist view by studying (i) how single neurons or small functional units regulate specific animal behaviors \cite{Cande2018}, or (ii) how simple interaction rules between self-propelled particles give rise to collective behaviors \cite{Vicsek1995,Toner1995,Couzin2011}. However, recent work is beginning to reveal that animal behavior---be it the movements of individuals or the foraging of groups---may not be fully explainable from the dynamics of individual constituent units \autocite{Jonas2017,Misic2016,Braganza2018,Chen2019} \textbf{(\autoref{figure_1})}. Hence a paradigm shift is needed to move from reductionist analyses to those that embrace the complexity of distributed information processing over networks spanning multiple levels \autocite{Baez-Mendoza2021,Rose2021}.

This shift is made possible thanks to technical advances that enable the acquisition of large-scale, high-resolution datasets. In neuroscience, connectomes and neural activity recordings have kickstarted the nascent field of network neuroscience \autocite{Bassett2017,Castro2020} which is beginning to draw connections between the structure and function of large-scale neural circuits. In collective behavior, the tracking of large animal groups has enabled the construction of inter-animal interaction networks \autocite{Kulahci2019,Firth2020}. A common feature of these datasets is that they can be represented as dynamic signals on the nodes of a network, such as neural activity or movement direction. These are amenable to modeling using network theory and graph signal processing \autocite{Ortega2018}.

Network theory is uniquely suited to form novel links across the neuroscience of individual and collective animal behaviors because it provides tools to discover universal principles of network dynamics that are robust to uncharacterised interaction parameters. In this review, we first describe new large-scale datasets that can be characterized by different network objects. Then we highlight four network theoretic concepts---signal propagation, controllability, symmetry and geometry---that have successfully linked network structure and function in the context of animal behavior and thus have the potential to generally influence thinking in the data-driven age.

\section*{Large-scale measurements of neural and animal interaction networks}

Here we review three kinds of datasets that are amenable to modeling over networks and consider their possible network definitions, including nodal and edge quantities.

\paragraph{Connectomic reconstruction of neural networks.}

\begin{figure*}[t]
    \centering
    \includegraphics[width = \textwidth]{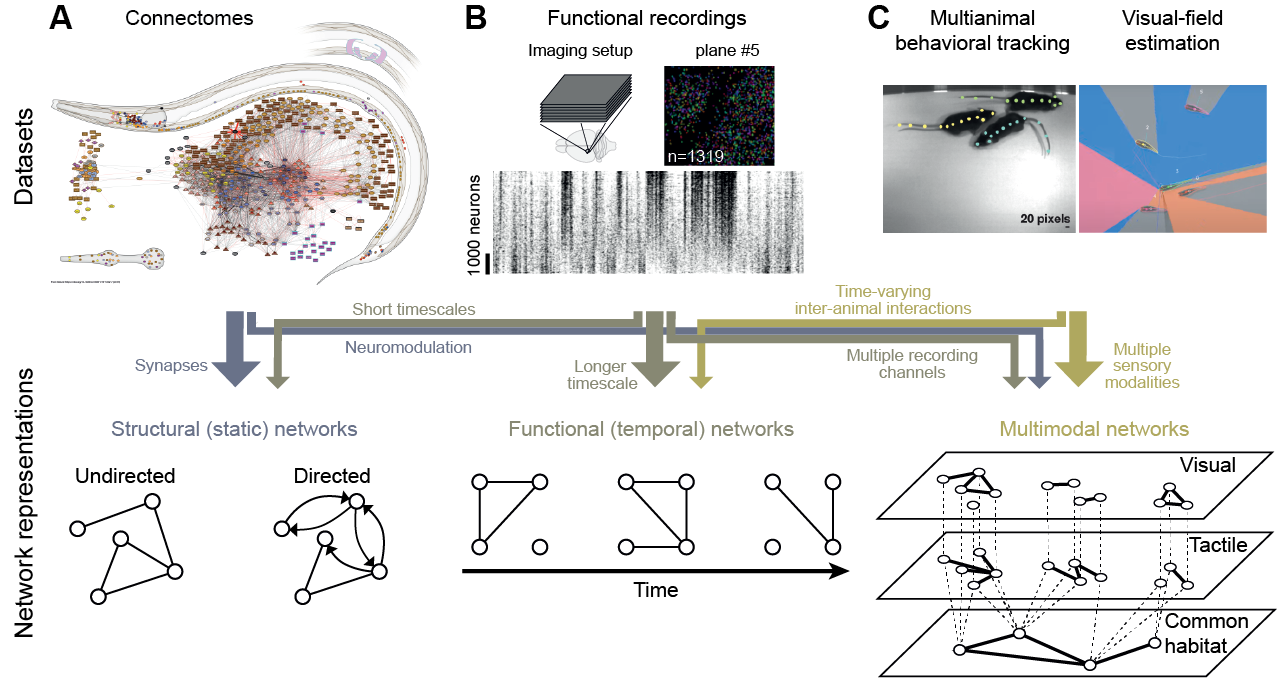}
    \caption{\textbf{Network descriptions of structural and functional data capturing the properties of neural and animal ensembles.}
    \textbf{(A)} Connectomes describe anatomical connections of neurons or neural compartments obtained using synaptic-resolution electron microscopy reconstruction (image of \textit{C. elegans} reproduced from \autocite{Cook2019}). Connectomes are typically represented as static networks in contexts where synaptic plasticity is irrelevant (e.g., short timescales). Multilayer networks can represent interactions between different classes of neurons.
    \textbf{(B)} Functional recordings reveal the dynamic activity of neural populations (image reproduced from \autocite{stringer2019b}). Functional networks encode a certain similarity between nodal dynamics. Because these connections covary with node dynamics, they can be represented as temporal networks, which in their simplest form are an ensemble of temporal network snapshots.
    \textbf{(C)} Animal tracking and visual field reconstructions can characterize behaviors: the output of neural activity. Thus, they provide a bridge between individual and collective animal behaviors (images reproduced from \autocite{Walter2021,Lauer2021}). Inter-animal interactions can be represented using functional networks. In addition, multilayer networks can represent multiple modalities not captured by a single network layer.
    }
    \label{figure_2}
\end{figure*}

A `connectome' is an extensive anatomical reconstruction of neural connections---typically through semi-automated segmentation of serial section electron microscopy data \textbf{(\autoref{figure_2}, left)}. The first connectome was obtained for the hermaphrodite sex of the worm \textit{C. elegans}, a resource that now includes nearly the complete nervous systems of both sexes (hermaphrodite and male) across development \autocite{Byrd2021}. Similar reconstruction efforts are underway for the \textit{Drosophila} central brain \autocite{Scheffer2020}, central complex \autocite{Hulse2020}, olfactory system \autocite{Schlegel2021}, and motor circuits in the ventral nerve cord \autocite{Phelps2021}. Beyond invertebrates, connectomics datasets have been generated for a larval zebrafish \autocite{hildebrand2017}, parts of mouse visual cortex \autocite{lee2016anatomy}, and human cerebral cortex \autocite{Shapson-Coe2021}.

Connectomes provide a structural network: individual neurons (nodes) connect to one another via directed chemical, or undirected/bidirectional electrical synapses (edges). Moreover, at a finer scale, network nodes may also represent dendritic compartments as fundamental units of computation \cite{Chavlis2021}. Because connectomic reconstructions typically involve one or at most a few network instances, their networks are generally considered static. This may be accurate on the timescale of animal behaviors, except when learning and plasticity occur. Edges are typically unweighted (i.e., having unit weight), but anatomical features of axons like their diameters have sometimes been used as a proxy for edge weights \autocite{Byrd2021}. This classical network model may be limiting when the heterogeneity of synaptic interactions plays an important role in network dynamics. In this case, `multilayer networks' can be used to account for different network features \autocite{DeDomenico2016,Finn2019}. Here, layers represent different modalities combined into a single mathematical object via inter-layer edges. For example, connections mediated by neuromodulators have been modeled as different network layers \autocite{Maertens2021}.

\paragraph{Functional recordings of large-scale neural networks.}

Complementing structural neuronal connectivity, optical functional recordings enable a readout of neural activity \textbf{(\autoref{figure_2}, middle)}. Although these recordings offer a lower temporal resolution than multi-electrode array recordings \autocite{steinmetz2021}, state-of-the-art genetic reagents enable the measurement of calcium influx \autocite{dana2019,zhang2020}, voltage \autocite{Jin2012,Villette2019}, or neuromodulator dynamics \cite{sabatini2020imaging} across large swaths of neural tissue, while also more effectively conveying information about each cell's type, identity, and spatial location. These functional datasets exist for a variety of small transgenic animals including \textit{C. elegans} \autocite{Randi2020,susoy2021}, larval \autocite{karagyozov2018} and adult \textit{Drosophila} \autocite{Mann2017,Chen2018,aimon2019fast,schaffer2021,Hermans2021}, larval zebrafish \autocite{cong2017}, and rodents \autocite{musall2019,stringer2019b,rumyantsev2020}.

Functional recordings represent dynamic signals over network nodes. These can be used to build a `functional network'. In this case, edge weights are not based on physical connections, but on a correlational or causative links between nodal dynamics. When edge weights represent correlational links, or `dynamic similarity', they typically covary with the node dynamics. Thus, functional networks are termed `temporal' in network theory, which at its simplest can be visualised as a multilayer network with layers encoding a sequence of discrete temporal snapshots \autocite{Li2017}.

A crucial current effort aims to synthesize functional and anatomical/connectomic datasets. This requires the mapping of cell identity across datasets, a challenging endeavour due to inter-animal variability in cell location as well as movement-related microscope image deformations. Progress on this front has been mostly limited to studies of \textit{C. elegans}, an animal for which the positions and identities of neurons are largely conserved across individuals. This fact facilitates multicolor labeling strategies to recover each neuron's identity from its spatial position and fluorescent protein expression profile \autocite{Yemini2021,Toyoshima2020}. Because this technique is not easily compatible with freely moving worms, alternative, deep learning-based methods have also been used to recognize and track neuron positions and identities across time \autocite{Chentao2021,Yu2021}.

\paragraph{Behavioral-tracking of animal collectives.}

A simple way to capture inter-animal interactions is by tracking their body positions in 2D space and using proximity as a readout of interactions \cite{Ramdya2015}. In addition, finer-scale 3D body kinematics can be precisely measured using deep learning-based markerless motion capture and multi-camera triangulation of multiple 2D poses \cite{Mathis2018,Gunel2019,Pereira2020}, or by lifting single-camera 2D poses \cite{Gosztolai2021b}. Recent methods have extended 2D pose estimation to multi-animal settings allowing investigators to track kinematics for up to $\sim$10 animals at once \autocite{Segalin2020,Pereira2020,Chen2020,Lauer2021} \textbf{(\autoref{figure_2}, right)}. Further insights may be gained by combining positional tracking of animals with body and head orientation measurements to infer their visual fields \autocite{Harpaz2021, Walter2021}.

Animals (nodes) and their pairwise interactions (edges) like spatiotemporal proximity, shared group membership, or behavioral similarity must be modeled as temporal networks \textbf{(\autoref{figure_2}, right)}. Furthermore, multilayer networks can account for the same individuals interacting via different sensory modalities, or individuals interacting across different spatial compartments \autocite{Finn2019}.

\section*{Applications of network theory for studying neural and animal social network dynamics}

\begin{figure*}[t]
    \centering
    \includegraphics[width = \textwidth]{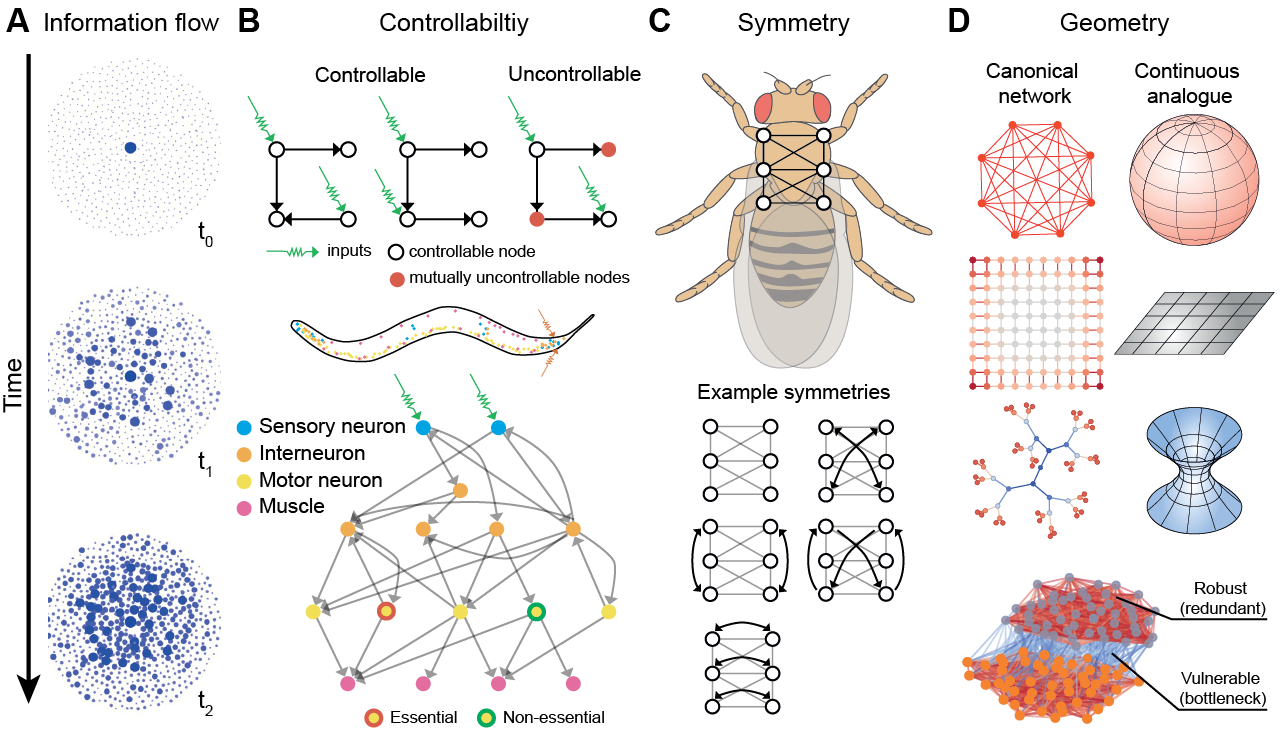}
    \caption{\textbf{Relevant network theoretical concepts for understanding neural and animal social network dynamics} 
    \textbf{(A)} Network dynamics can be conceptualized as signals evolving on network nodes (edges are not shown; image reproduced from \autocite{Hens2019}).
    \textbf{(B)} Controllability of a network quantifies the role of inputs in influencing network dynamics. The networks on the left are controllable because each node can be driven to an arbitrary state by the two inputs. The network on the right is uncontrollable because the red nodes cannot be driven to arbitrary states by the inputs. Controllability has been used to identify \textit{C. elegans} sensory neurons that affect the global network state and thus are of likely behavioral relevance (schematic based on \autocite{Yan2017}).
    \textbf{(C)} Network symmetries are closely related to permissible network dynamics. The function of large networks can be decomposed into small network units or motifs with well-characterised symmetries and input-output relationships. Here, the symmeties of a network of six central pattern generators (CPGs) controlling insect locomotion can provide insights into functional circuit dynamics generating gaits.
    \textbf{(D)} Network geometry aims to represent network structure as a mathematical object that can better reveal network symmetries or dynamical properties. One example is edge curvature, which is typically defined based on some analogy to canonical geometries such as balls, planes or hyperboloids. Amongst other predictions, network geometry can infer a network's robustness (redundancy), or vulnerability (bottlenecks).}
    \label{figure_3}
\end{figure*}


\paragraph{Modeling signal propagation through a network.}

The behaviors of neural and animal collectives can be thought of as dynamical signals propagating along the edges of network nodes. This is often referred to as dynamic flow or information spreading in statistical physics \autocite{Harush2017} \textbf{(\autoref{figure_3}A)}. Indeed, the activity of a node influences the likelihood that a neighboring node changes its activity which can lead to a cascade propagating throughout the network.
A model-based dynamical systems approach to understand the patterns of this flow entails considering nodes as state variables (e.g., the firing rate of a neuron, or swimming velocity of a fish) coupled through often non-linear interactions. The interaction model, which can be phenomenological (e.g., Integrate-and-Fire neurons) or mechanistic (e.g., Hodgkin-Huxley neurons), explicitly describes how state variables change as a function of other variables. This framework typically entails suitably parametrising the interaction functions and performing simulations of the network's activity \autocite{Randi2020}. Although this approach is useful for making predictions about network activity, it is unlikely to provide generalizable insights because there is no one-to-one mapping between dynamics and parameters. For example, neural networks can exhibit the same dynamics despite morphological variations of neurons, heterogeneous circuit parameters, and neuromodulation \autocite{Katz2016,marder2015,Ji2020}, and, conversely, networks can support different dynamics despite very small variations in connectivity \autocite{Harush2017}. 

Network theory provides an alternative approach that is inspired by epidemic or rumor-spreading models in which nodes adopt their neighbors' states---such as whether they are susceptible or infected. Likewise, neurons change their activity depending on the states of neighboring neurons, the interaction rules between them, and their processing at nodes. Similarly, for animal collectives, epidemic models capture the process whereby changes in an individual's behavior propagates through the network \autocite{centola2018}. This abstraction is powerful because it allows one to translate a network's topology to observed signal propagation patterns without requiring a detailed characterisation of dynamical interactions.

To understand the role of individual neurons in network-wide signal propagation, a common simplifying assumption is to model network activity as a linear process. In \textit{C. elegans}, this technique has been effective to predict which nodes (neurons), when removed, cause maximal disturbance in flow patterns \autocite{bacik2016}. This is likely because the worm's nervous system consists of many neurons communicating using gap junctions, which can be modeled as linear resistors. Although chemical synapses may introduce nonlinearities, their sigmoidal transfer functions are well-approximated by a linearization around their operating point \autocite{bacik2016}. However, this linear approximation may also generally apply to other organisms because nonlinear neural dynamics often evolve on a low-dimensional manifold \autocite{Gallego2017,Chung2021} that is also well-approximated by linearization in the neighborhood of a point in neural space. Taking advantage of this feature, one study examined the dynamics of a linearized system and formed a new `similarity' network where edge weights represent pairwise correlations between nodal dynamics \autocite{Schaub2018}. Clustering this network predicted which groups of neurons were likely to be coactive in the nonlinear system. It is known that signal propagation patterns depend on nonlinear node dynamics \autocite{Harush2017}. Yet, strikingly, for a variety of networks in neuroscience, ecology, and epidemiology, spreading behaviors fall into distinct modes depending on purely structural features. These include the shortest paths between nodes and high degree nodes (hubs) \autocite{Hens2019}. Studies aiming to understand the global effect of nonlinearities are extremely valuable for predicting how specific features of neural tuning can influence large-scale network computations.

Epidemics-inspired models are also insightful in the study of animal collectives. Indeed, animal interactions typically depend only on the relative position of individuals except, for example, in cases of crowding \autocite{Davidson2021}. Modeling the behavioral changes mediated by network interactions, often termed social contagion, is simpler than modeling the evolution of a population's state, which has traditionally been studied using approaches from the kinetic theory of gases \autocite{bodova2018}. Early `simple contagion' models considered the probability of an individual adopting a new behavior to be proportional to the number of neighbors with that behavior \autocite{hoppitt2017}. However, it is now recognized that this probability must include a nonlinear function of neighboring behaviors, known as a `complex contagion' \autocite{Firth2020}. For example, in schooling fish only models accounting for the cooperative effects of neighboring active individuals can explain group dynamics \autocite{Rosenthal2015}. Thus, by simulating social cascades it has been possible to distinguish the effect of individual-level parameters from that of the group's structure \autocite{Sosna2019}. The dynamics of signal propagation have also been extended to multilayer networks to reveal the roles of different interaction modes \autocite{DeDomenico2016}.

\paragraph{Controllability of network signals.}

In addition to network structure, network dynamics are also shaped by inputs \autocite{Kunert2017}, such as sensory inputs driving neural networks, or predators disturbing animal interaction networks. Inputs can affect the network locally, or they can spread to the majority of nodes. `Controllability'  is the notion that links network signals to their inputs \autocite{Schaub2018}. This measures the ability of an input to drive network states to a desired target in finite time \autocite{Kao2019}. A special case that assumes a linear system is `structural controllability', which tests whether an input to a specific node can significantly affect network dynamics. Although it is a linear property, structural controllability predicts the minimal set of inputs sufficient to control an underlying nonlinear system \textbf{(\autoref{figure_3}B)}. When applied to connectomes, structural controllability can infer which inputs---from sensory organs, or other brain regions---are behaviorally relevant. Classically, this task has been performed by ablating sensory neurons and subsequently searching for a loss of function: an approach that is experimentally intractable for larger groups of neurons or to probe combinations of neurons. One study examined structural controllability of the \textit{C. elegans} connectome to predict sensory neuron classes, as well as single neurons within these classes, whose removal would reduce the number of controllable muscles, thus impairing locomotion \autocite{Yan2017} \textbf{(\autoref{figure_3}B)}. 

Controllability has also been generalized to temporal networks, which are useful for studying animal collectives \autocite{Li2017}. In temporal networks, signal propagation is slower but control is easier because the increased number of layers enlarges the space of possible control trajectories \autocite{Li2017}. For example, controllability can be achieved more rapidly in a network of antenna-body interactions in ants, compared to a network composed only of static interactions \autocite{Li2017}. Thus, control theory can provide insights into how brains generate robust actions while also enabling diversity at the level of individual and group-level behaviors.

\paragraph{Understanding dynamics through network symmetries.}

Networks may contain `structural symmetries'. These are possible rearrangements of nodes that leave network topology invariant or unchanged \textbf{(\autoref{figure_3}C)}. In neural networks, structural symmetries are required for controllability \autocite{Whalen2015} and for synchronisation \autocite{golubitsky2012}. For example, central pattern generators, which are frequently used to model animal locomotion, must have ipsi- and contralateral symmetries to generate locomotor gaits \autocite{stewart2017}. Based on this insight, one study suggested that certain \textit{C. elegans} locomotor patterns are associated with structural symmetries in the worm's connectome \autocite{Morone2019}. They found that circuits regulating forward/backward locomotion can be decomposed into a hierarchical system of dynamical units (filters) with well-defined symmetries. The dynamics of these units contribute to locomotion but are largely independent of the specific dynamic parameters of the neurons. This decomposition is related to network motifs---network units with well-characterised input-output behaviors \autocite{Braganza2018}. As a result, evidence for network motifs between pairs of nodes can also be found experimentally by injecting a prescribed dynamic signal into one node and looking for certain dynamical signatures in other nodes \autocite{Rahi2017}. Similar symmetry-function relationships are also emerging for animals with larger nervous systems. For example, the connectivity, inputs, and outputs of the \textit{Drosophila} central complex have recently been examined to link circuit motifs with potential functional properties \autocite{Hulse2020}.

\paragraph{Network geometry linking dynamics and structure.}

Further links between network symmetries, controllability and signal propagation can be discovered using tools from the emerging field of network geometry \autocite{Boguna2021}. Briefly, network geometry aims to represent a network by either identifying a continuous latent space in which it can be embedded or by defining a geometric object based on features of the network's structure or node signals. The motivation behind constructing geometric objects is that they may be particularly suited to reveal structure-function relationships. For example, geometric notions have been exceptionally useful in identifying hidden symmetries and predicting the spatiotemporal evolution of network-driven dynamical processes \autocite{Boguna2021}. A geometric approach has been used to uncover symmetries in the human functional connectome, suggesting universal organizational principles across scales \autocite{Muhua2020}. Network geometry has also been used to infer information-limiting bottlenecks between regions \cite{Gosztolai2021} and those that are redundant for signal propagation \cite{Farooq2019,Gosztolai2021}. Structural features like association to a high-degree node might not highlight these properties. Thus, network geometry has the potential to predict the relevance of connections from dynamic network models or neural recordings. Network geometric ideas have also been used in collective behavior to detect dynamical transitions when a hidden parameter is varied. In \autocite{Skinner2021}, the authors noticed that, in a collective system, the state of the whole system can be encoded as a probability distribution over the local connectivitiy of each individual. Thus, they could compare the dynamics across different conditions based on their respective probability distributions. Using this approach, they were able to detect dynamical transitions in collective behavior without temporal information but based purely on changes in the relative arrangement of individuals.

\section*{Conclusions}
Recent technical advances have enabled the acquisition of large-scale datasets in neuroscience and collective behavior. These can be represented as networks of neural connectivity, functional dynamics, and also population-level inter-animal interactions. Network theory offers a set of tools that can help to distill universal principles from these data, linking structure and function, often from only a few noisy network instances. Progress in this direction will offer new avenues for investigating distributed computations performed by collective systems of neurons and animals, and can advance machine learning approaches that leverage the power of bioinspired network operations.

\section*{Competing interests}
Declaration of interest: none.

\section*{Acknowledgements}
The authors thank the Neuroengineering lab and in particular Jonas Braun, Gizem \"{O}zdil, and Matthias Durrieu for insightful comments on the manuscript.
PR acknowledges support from an SNSF Project grant (175667), and an SNSF Eccellenza grant (181239). AG acknowledges support from an HFSP Cross-disciplinary Postdoctoral Fellowship (LT000669/2020-C).

\clearpage

\defbibnote{myprenote}{*Papers of special interest\\
* **Papers of outstanding interest}
\printbibliography[prenote=myprenote]

\end{document}